\documentclass[reprint,aps,pre,onecolumn,notitlepage,superscriptaddress,longbibliography]{revtex4-1}
\usepackage{graphicx}
\usepackage{amsmath,amssymb}
\usepackage{float}
\usepackage{comment}
\usepackage{bm}
\usepackage{color}
\usepackage{dcolumn}
\usepackage{siunitx}
\newcolumntype{x}[1]{>{\centering\arraybackslash\hspace{0pt}}p{#1}}
\newcommand{\reffig}[1]{Fig.~\ref{#1}}

\newcommand{\OSP}{OCS} 

\newcommand{\subsec}[1]{\noindent\textbf{\textcolor{black}{\\#1\\}}}

\newcommand{\SSrate}{\dot{\gamma}}
\newcommand{\OSPamp}{\gamma^{\mathrm{}}}
\newcommand{\GSOamp}{\gamma^{\mathrm{}}}
\newcommand{\OSPfreq}{\omega^{}}
\newcommand{\OSPfreqND}{\OSPfreq\OSPamp/\SSrate}

\begin{document}

\preprint{APS/123-QED}
\title{Shaken \emph{and} stirred: Random organization reduces viscosity and dissipation in granular suspensions}
\author{Christopher Ness}
\affiliation{Department of Chemical Engineering and Biotechnology, University of Cambridge, Cambridge CB3 0AS, United Kingdom}
\author{Romain Mari}
\affiliation{DAMTP, Centre for Mathematical Sciences, University of Cambridge, Cambridge CB3 0WA, United Kingdom}
\affiliation{Universit\'e Grenoble Alpes, CNRS, LIPhy, 38000 Grenoble, France}
\author{Michael E. Cates}
\affiliation{DAMTP, Centre for Mathematical Sciences, University of Cambridge, Cambridge CB3 0WA, United Kingdom}
\date{\today}

\begin{abstract}
The viscosity of suspensions of large ($\geq\SI{10}{\micro\meter}$) particles diverges
 at high solid fractions due to proliferation of frictional particle contacts.
 Reducing friction, to allow or improve flowability, is usually achieved
 by tuning the composition, either changing particle sizes and shapes or by adding lubricating molecules.
We present numerical simulations that demonstrate a complementary approach
 whereby the viscosity divergence is shifted by \emph{driven flow tuning},
 using superimposed shear oscillations in various configurations to facilitate a primary flow.
The oscillations drive the suspension towards an out-of-equilibrium, absorbing state phase transition,
where frictional particle contacts that dominate the viscosity are reduced in a self-organizing manner.
 The method can allow otherwise jammed states to flow; even for unjammed states, it can substantially decrease the energy dissipated per unit strain. This creates a practicable route to flow enhancement across a broad range of suspensions
 where compositional tuning is undesirable or problematic.

 \end{abstract}

\maketitle

\section*{Introduction}
Densely packed suspensions arise widely in industry and manufacturing,
where reliable, predictable and prescribable flow properties are essential~\cite{benbow1993paste}.
A major limiting factor in their processability is
the very steep increase of viscosity upon increasing the volume fraction of solid material $\phi$
towards the jamming transition~\cite{liu1998nonlinear}.
This is particularly evident in the non-Brownian regime (particle size $\geq \SI{10}{\micro\meter}$)
where \emph{frictional} particle contact interactions reduce the jamming density $\phi_m$~\cite{silbert_jamming_2010,song2008phase,Guy2015}
and increase dissipation in process flows, resulting in high energy costs.

Empirical strategies that reduce the viscosity and/or net dissipation
include tuning the physical properties of the particles ---for example their size, shape and polydispersity---
or modifying their interactions through chemical additives known as plasticizers, emulsifiers or friction modifiers. These lubricate interparticle contacts and reduce the suspension viscosity by raising $\phi_m$.
Often, though, end-use requirements leave little room for manoeuvre in the formulation.
In calcium phosphate cements for bone injection~\cite{o2016extent,zhang2014calcium}, for example,
chemistry and biology both constrain the use of molecular additives.
There is therefore a practical need for methods of dense suspension flow control that do not require changes to formulation.

Two recent experiments suggest a possible route towards this goal, achieving \emph{driven} viscosity reduction
by superimposing an oscillatory cross shear (\OSP) on a primary desired flow.
Using \OSP, Blanc~\emph{et al}~\cite{blanc_tunable_2014} demonstrated
a two-fold increase in the sedimentation velocity of an intruder in
a granular suspension of rate-independent rheology,
while Lin~\emph{et al}~\cite{Lin2016a} measured a two decade viscosity drop in
one of shear-thickening rheology. The latter effect was argued to be a consequence of the fragility
of shear-induced particle contacts~\cite{Cates1998,Lin2016a}, suggesting that
good flowability might be achieved only when the \OSP{} is sufficiently fast to keep
the microstructure in a load-incompatible state~\cite{Cates1998,lin2015hydrodynamic}.
In this limit, the reduction in primary flow viscosity (unless this is infinite) might easily be outweighed
by the high energy cost of implementing fast \OSP{}, particularly for rate-independent suspensions~\cite{blanc_tunable_2014} whose primary viscosity drop is much less than in shear-thickening ones~\cite{Lin2016a}.
More generally,
it is not clear how far the benefits of \OSP{}
depend on the underlying suspension rheology: the short-range repulsions
that prevent frictional particle contacts at low stresses in thickening suspensions~\cite{seto_discontinuous_2013,clavaud_revealing_2017,comtet2017pairwise}
may or may not play a major role during \OSP{}-assisted viscosity reduction.

In this article, we present numerical simulations showing that
the viscosity drop induced by \OSP{} is
generic to suspensions with friction-dominated stress.
This includes noninertial flows of most dense suspensions of super-micron sized particles~\cite{Guy2015}. 
The transverse flow oscillations directly inhibit particle contacts without requiring short-range repulsions,
enhancing lubrication and shifting $\phi_m$ to higher values.
Consequently, the viscosity reduction increases with increasing $\phi$,
so that near jamming the saving in primary flow dissipation outweighs
the cost of \OSP{} at any primary flow rate, giving a net reduction in the energy expended per unit strain in the primary direction.
We then show that the reduction in particle contacts stems from an \OSP-induced `random organization' mechanism
~\cite{pine_chaos_2005,corte_random_2008,corte_self-organized_2009}; this leads us to
an enhanced version of the flow protocol that can reduce the dissipation further.
Guided by these results, we argue that driven viscosity control should extend flowability and reduce the associated energy cost
across a broad class of materials including slurries, muds, cement and other immersed granular systems.

\begin{figure*}
 \includegraphics[trim={0mm 25mm 51mm 0mm},clip,width=1\linewidth]{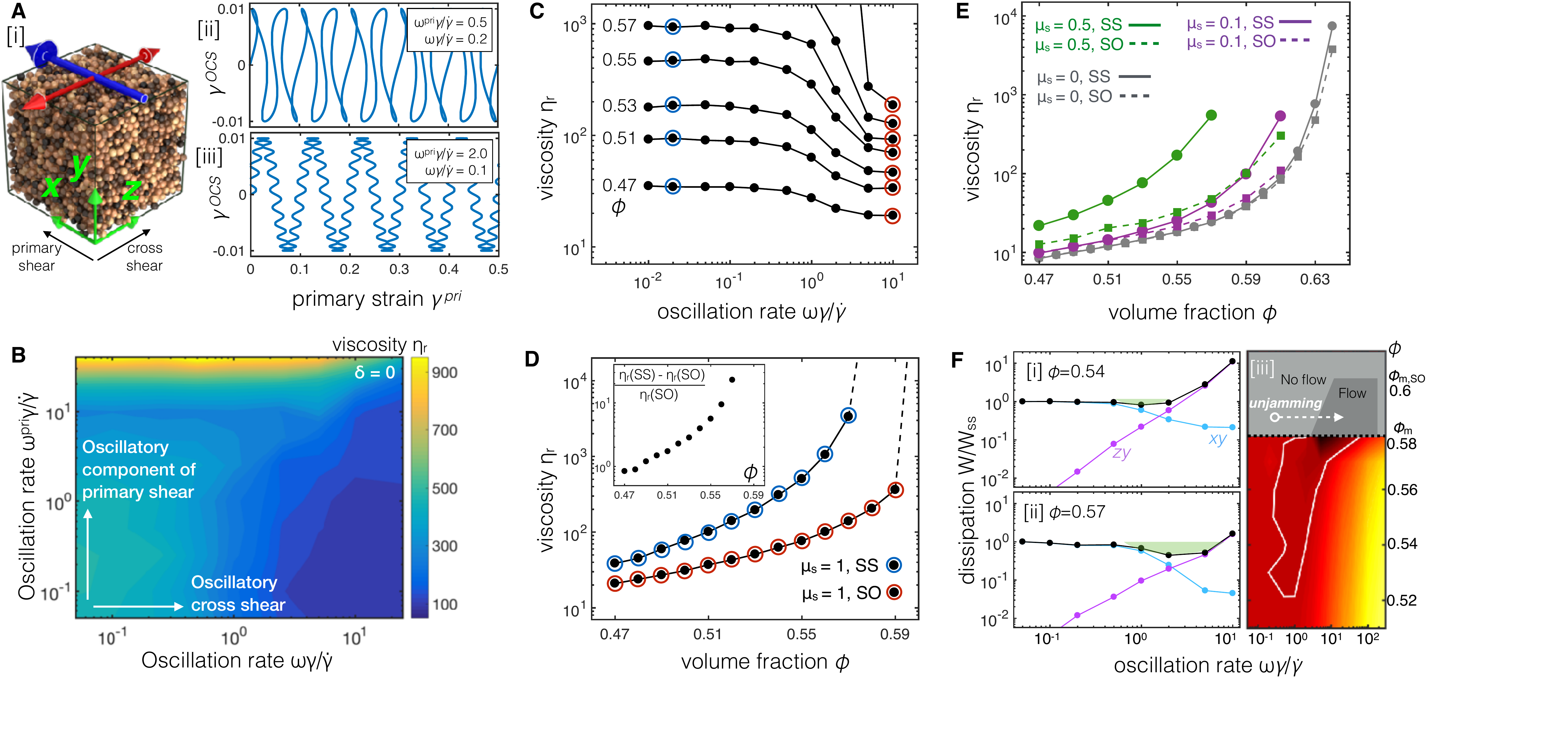}
 \caption{
  Viscosity and dissipation reduction under superimposed primary and oscillatory flows.
  \textbf{A}:
  [i] Simulation snapshot showing primary (blue) and cross shear (red) flow directions;
  [ii] and [iii] Example flow paths explored for different values of parameters $\omega^\mathrm{pri}$ and $\OSPfreq$ (values given in Insets) with $\delta=0$;
  \textbf{B}: Contour map showing viscosity in primary flow direction as a function of $\omega^\mathrm{pri}$ and $\OSPfreq$, for $\delta=0$ and $\phi=0.55$;
  \textbf{C}: Viscosity as a function of oscillation rate $\OSPfreq{\gamma}/\dot{\gamma}$ at various volume fractions $\phi$ with amplitude $\OSPamp= \SI{1}{\percent}$ and $\omega^\mathrm{pri}=0$, the simple OCS (SO) protocol;
  \textbf{D}: Viscosity divergence as a function of $\phi$ under steady shear (SS) and high frequency SO with friction coefficient $\mu_s=1$. Inset: difference between SS and SO viscosities;
  \textbf{E}: Viscosity divergences for particles with lower friction coefficient $\mu_s$ show diminishing viscosity reduction;
  \textbf{F}: Dissipation per unit strain $W$ (rescaled by the $\phi$-dependent steady shear dissipation $W_\text{SS}$) as a function of oscillation rate for the same simulations as in \textbf{C}, for [i] $\phi=0.54$ and [ii] $\phi=0.57$ showing contributions in $xy$ and $zy$.
  Green areas in [i] and [ii] highlight the region in which both viscosity and dissipation reduction are achieved.
  [iii] Dissipation in the ($\OSPfreq\OSPamp/\SSrate$, $\phi$) plane, highlighting in white the region for which dissipation may be reduced by at least 5\% with SO.
 }
 \label{figure1}
\end{figure*}

\section*{Results and discussion}
We study a suspension of nearly-hard, athermal spheres
subject to short-range hydrodynamic and contact interactions with static friction coefficient $\mu_s$ as described in Methods below.
This numerical model (and similar ones~\cite{gallier2014rheology,peters2016rheology}) is known to yield accurate predictions for the rheology of
non-Brownian hard sphere suspensions.
The suspension shows rate-independent rheology,
well described under steady simple shear by the viscous number formalism~(see \cite{boyer2011unifying} and SI).
A snapshot of the simulated system is shown in Fig~\ref{figure1}A[i].

\subsec{Manipulating suspension viscosity using superimposed oscillations}
From the argument that fragility makes contact stresses in suspensions susceptible to driven perturbations~\cite{Lin2016a},
it follows that the addition of \emph{any} arbitrary oscillating flow might lead to viscosity reduction.
This hypothesis is in line with experimental~\cite{janda2009unjamming,hanotin2012vibration} and theoretical~\cite{degiuli2017friction} works
that propose applied and endogenous noise, respectively,
as sources of opening and closing granular contacts and consequent unjamming.
To test this, we first explore a generalization of \OSP{} comprising primary steady shear with rate $\SSrate$ and superimposed oscillatory shears in {\em both} the primary and cross shear directions,
leading to an overall strain in $xy$ as
${\gamma}^\mathrm{pri}(t) = \GSOamp \sin(\omega^\mathrm{pri} t+\delta) + \SSrate t\text{,}$
and in $zy$ as
${\gamma}^\mathrm{\OSP}(t) = \GSOamp \sin(\OSPfreq t) \text{.}$ For simplicity we
keep $\GSOamp=1\%$ in each case, which~Ref~\cite{Lin2016a} found to be an optimal amplitude for viscosity reduction.
The remaining dimensionless control parameters are then $\omega^\mathrm{pri}\GSOamp/\SSrate$, $\OSPfreq\GSOamp/\SSrate$ and
the phase shift $\delta$.
This protocol gives strain paths such as those illustrated in~Figs.~\ref{figure1}A[ii]~and~[iii].
A characteristic viscosity is computed as $\eta_r=\sigma_{xy}/\eta\SSrate$ averaged over $\sim10/\SSrate$ time units,
with $\eta$ the solvent viscosity and $\sigma_{xy}$ the $xy$ component of the stress.
We find that $\delta$ has very little effect on the viscosity (see SI),
and present a contour map of $\eta_r$ in the
($\omega^\mathrm{pri}\GSOamp/\SSrate$, $\OSPfreq\GSOamp/\SSrate$)
plane at $\delta=0$ and $\phi=0.55$ in~Fig.~\ref{figure1}B.
At fixed $\delta$, viscosity minima are obtained as
$\omega^\mathrm{pri}\GSOamp/\SSrate\to0$
and
$\OSPfreq\GSOamp/\SSrate\gtrsim6$.
In this limit, i.e. with cross shear oscillations only, we obtain a viscosity drop
comparable to that for a thickening suspension~\cite{Lin2016a}.
This suggests that \OSP{} -- by keeping frictional particle contacts open -- effectively brings the suspension to a low-friction state. 
We similarly find a maximal rate of viscosity reduction when
$\OSPfreq\GSOamp/\SSrate$ is close to unity. Contrary to the hypothesis made above, however, our results show that the orientation of the oscillatory flow is crucial: at this strain amplitude, any oscillatory component {\em along} the primary flow direction makes no useful contribution to improving flowability.
If the shear is constrained to a single direction and the amplitude of the oscillations is very small compared to the primary flow, the net displacements of the particles over large strains are, for rate-independent flow, the same as for steady shear. This is not the case when the oscillations are applied transverse to the primary flow.

\begin{figure*}
  \includegraphics[trim={0mm 139mm 0mm 0mm},clip,width=1\linewidth]{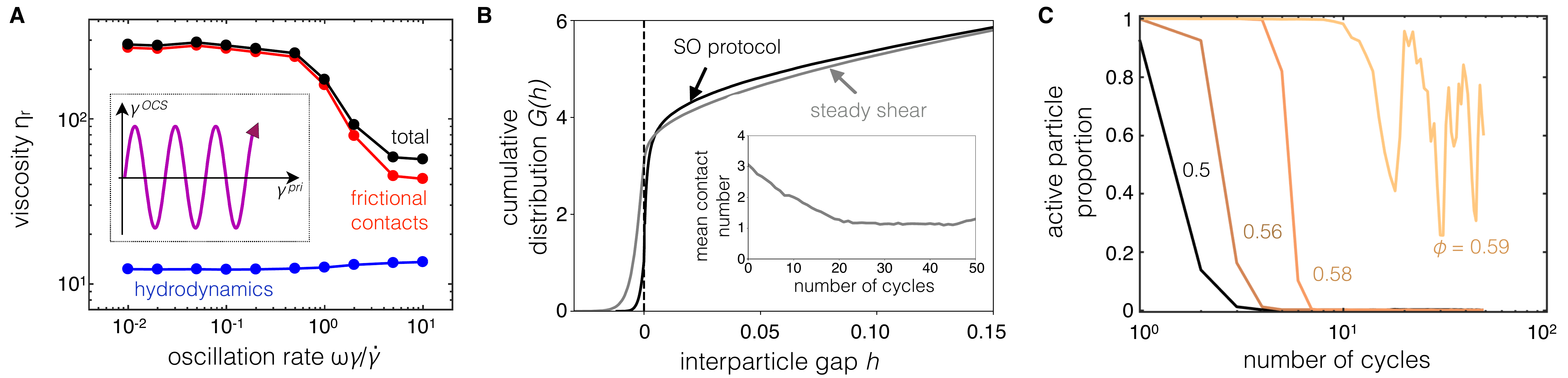}
  \caption{
   Revealing random-organization at work during oscillatory shear.
   \textbf{A}: Origin of the viscosity drop at $\phi=0.54$ with SO. The contact stress contribution is strongly suppressed with increasing oscillation rate. Inset: schematic of the SO strain profile;
   \textbf{B}: The cumulative pair correlation function $G(h)$ under steady shear and $\OSPfreqND=10$, demonstrating a room-making process;
   Inset: Around 20 cycles are needed to minimize the number of particle contacts
   \textbf{C}: Proportion of particles following an irreversible trajectory under successive periods of oscillatory shear (in the absence of primary shear) as a function
   of the number of cycles, starting from presheared configurations at several volume fractions.
   The steady decrease of the irreversibility is a signature of random organization.
  }
  \label{figure2}
\end{figure*}

\subsec{Simple OCS (SO): viscosity reduction using transverse oscillations}
In what follows we therefore revert to the purely transverse case with $\omega^\mathrm{pri}=0$, leading to $\gamma^\mathrm{pri}(t)=\dot{\gamma}t$,
and ${\gamma}^\mathrm{\OSP}(t) =  \OSPamp \sin(\OSPfreq t)$ (see~\reffig{figure2}A [Inset]),
hereafter called the ``simple \OSP{}'' protocol, SO. (This is to distinguish it from an alternative protocol introduced below.)
In Fig.~\ref{figure1}C we report the viscosity $\eta_{r}$
under this protocol at $\OSPamp=\SI{1}{\percent}$ as a function
of the reduced frequency $\OSPfreqND$, while in Fig.~\ref{figure1}D we compare,
as a function of volume fraction $\phi$,
the steady shear viscosity (obtained when $\OSPfreqND=0$) to the limiting viscosity under SO
(obtained when $\OSPfreqND\geq10$).
The viscosity drop increases rapidly with $\phi$,
reaching a decade at $\phi = 0.56$
and actually diverging between $\phi = 0.57-0.58$ (Fig.~\ref{figure1}D Inset).
This reveals that as well as reducing the viscosity,
the effect of SO is to slightly delay the jamming transition
from $\phi_m\approx0.58$
for steady shear to $\phi_{m,\mathrm{SO}}\approx0.60$
at $\OSPfreqND=10$.
Though small in absolute terms,
shifts of jamming by a couple of percent can have dramatic consequences for formulation and processing~\cite{benbow1987extrusion}, as discussed further below.

This shift of jamming under SO naturally raises the question
of the sensitivity of the viscosity reduction to particle friction,
which may stem from e.g. surface roughness~\cite{lootens_dilatant_2005}.
It is expected that in the absence of ordering, which we do not observe in our binary system,
the random close packing density $\phi_\mathrm{RCP}\approx 0.64$ is an upper limit for both
$\phi_m$ and $\phi_{m,\mathrm{SO}}$.
Furthermore, it is established that the jamming point $\phi_m$
approaches $\phi_\mathrm{RCP}$ as surface friction $\mu_s$ is decreased~\cite{chialvo2012bridging,hsiao2016rheological}.
As $\phi_m<\phi_{m,\mathrm{SO}}$, it follows that $\phi_m < \phi_{m,\mathrm{SO}} < \phi_\mathrm{RCP}$
and the window between $\phi_m$ and $\phi_{m,\mathrm{SO}}$ consequently vanishes
in the limit of low friction (as $\mu_s\to0$).
The performance of SO thus diminishes as friction decreases.
We demonstrate this in Fig.~\ref{figure1}E in the limits of steady shear
and SO with $\OSPfreqND=10$.
As a result, suspensions of rough particles, which are typically the most problematic in terms of processing~\cite{gallier2014rheology}, are best placed to benefit from driven flow control.
In the context of friction-driven shear thickening of colloids, this result thus confirms that SO can be successful in reducing the viscosity of a \emph{thickened} sample, as demonstrated by~Ref~\cite{Lin2016a}, but that it would fail to reduce the viscosity of a \emph{non-thickened sample}, i.e. one at which the applied stress lies below the onset stress~\cite{seto_discontinuous_2013,mari2014shear,wyart2014discontinuous,Guy2015}.


\subsec{SO-enabled reductions in energy dissipation}
The ability to control suspension viscosity during flow is itself desirable for mitigating instabilities~\cite{hermes2016unsteady} and, for example, when pumps are desired to operate within narrow bounds.
Often, though, rheological tuning has a somewhat different objective:
to minimize the energy cost of processing.
For $0.58<\phi<0.60$, oscillatory cross shear triumphs: it permits flow at finite dissipation rates not otherwise possible.
Below jamming ($\phi<\phi_m$) however, its benefits are less obvious.
The energy dissipation is given per unit volume and per unit primary strain as
$W = \lim_{T\to \infty} \left(\int_{0}^{T}\mathrm{d} t \bm{\sigma}:\bm{\dot{\gamma}}\right)/(\gamma^\text{pri}(T)-\gamma^\text{pri}(0))$.
Figs.~\ref{figure1}F[i]-[ii] show this quantity (rescaled by the $\phi-$dependent steady shear dissipation $W_\mathrm{SS}$) as a function of oscillation rate for our SO protocol,
separating out the primary ($\sigma_{xy}\SSrate$) and cross flow ($\sigma_{zy}\dot{\gamma}^\mathrm{\OSP}$) contributions.
The primary dissipation decreases in line with the viscosity, but
the direct cost of the cross shear increases as $(\OSPfreqND)^2$.
Summing these,
we identify oscillation rates $\OSPfreqND$ for which $W$ is usefully decreased,
highlighted green in Figs.~\ref{figure1}F[i]-[ii], and outlined in white in Fig.~\ref{figure1}F[iii]. This operating window, although it grows as the density approaches $\phi_m$, remains narrow at lower densities.
We show below that it can be extended significantly by a simple modification to the oscillatory protocol.

\subsec{Random organization drives the viscosity reduction}
The proposed modification exploits mechanistic insights, gleaned from our simulations, into how oscillatory cross shear promotes flowability.
To gain these insights,
we start by decomposing the viscosity into its hydrodynamic and frictional particle contact contributions,
revealing that at $\phi=0.54$ the stress is dominated by friction for any oscillatory frequency,~\reffig{figure2}A.
Significantly, the effect of the cross shear oscillations is to decrease this frictional part, while leaving the hydrodynamic
part unchanged.
The loss of friction parallels the shift of jamming to higher $\phi$ (\reffig{figure1}D),
indicative of a shift from rolling to sliding contacts as $\OSPfreqND$ is increased~\cite{wyart2014discontinuous,Guy2015}.
Defining interparticle gaps $h_{ij} = 2(r_{ij} - a_i - a_j)/(a_i+a_j)$
with $r_{ij}$ the centre-to-centre distance between particles $i$ and $j$
with radii $a_i$ and $a_j$ respectively,
we compute $G(h)$,
the average number of neighbours around a particle separated at most by $h$,~\reffig{figure2}B.
The loss of frictional particle contacts occurs by a `room-making' process,
whereby the mean distance between nearest neighbours increases.
Consequently, starting with a presheared sample there is a gradual decrease
of particle contacts over $\mathcal{O}(10)$ cycles after SO startup, \reffig{figure2}B [Inset].
Strikingly, room-making does \emph{just enough} to
hinder the stress-generating contacts.
This is strongly reminiscent of `random organization', whereby
application of oscillatory shear to a suspension of hard particles
drives collective self-organization, leading to configurations that minimize
the number of particle contacts generated per cycle~\cite{pine_chaos_2005,corte_random_2008,corte_self-organized_2009}.
Below a critical volume fraction, the system evolves to an `absorbing state'
for which configuration invariance is ensured
under further oscillations. Though first elaborated for dilute suspensions, a similar scenario applies at higher density where the absorbing state is defined not by absence of collisions but absence of plastic rearrangements~\cite{nagamanasa2014experimental,royer2015precisely}.

Flow-induced random organization offers a natural explanation for the viscosity decrease upon increasing oscillation frequency.
Indeed, at high frequency the `primary' and `secondary' labels respectively assigned
to steady shear and oscillatory cross shear are misnomers. In fact we have a steady transverse flow that weakly perturbs an oscillatory flow, for which the random organization effect is well established.
This makes room around particles, thus decreasing the contact stress,
while the steady shear slowly consumes this room and simultaneously initiates new particle contacts.
At finite `primary' flow, the absorbing state can never be reached but its proximity allows particles to avoid frictional particle contacts at densities where these would otherwise cause large viscosities or jamming.

The largest $\phi$ at which absorbing states are obtained locates
a nonequilibrium phase transition, where the self-organization process
is maximised~\cite{pine_chaos_2005,corte_random_2008,pham2016origin}.
To confirm the role of random organization,
we determine the location of this transition in our system.
Starting from a presheared configuration,
we apply an oscillatory shear
at $\OSPamp = \SI{1}{\percent}$ with no primary flow,
and measure the fraction of particles following
irreversible, `active', trajectories, which for these purposes we define as those whose net displacement after a cycle stays below a threshold of $10^{-5}a_i$.
For $\phi\leq0.58$, this quantity approaches zero,
indicating that the system evolves towards absorbing states,~\reffig{figure2}C.
For an amplitude $\OSPamp = \SI{1}{\percent}$, the absorbing state transition density is thus estimated as $0.58<\phi<0.59$.
Since in practice the transition is cut off by the primary flow, our precise definition of activity is not crucial here, although a more inclusive one (e.g., counting all particles that make frictional contact at any point during the cycle)
would give a lower estimate for the transition. Nonetheless, our results indicate that the random organization effect is indeed a strong one throughout the density range where our SO protocol is effective.

\begin{figure*}
 \includegraphics[trim={0mm 8mm 0mm 0mm},clip,width=1\linewidth]{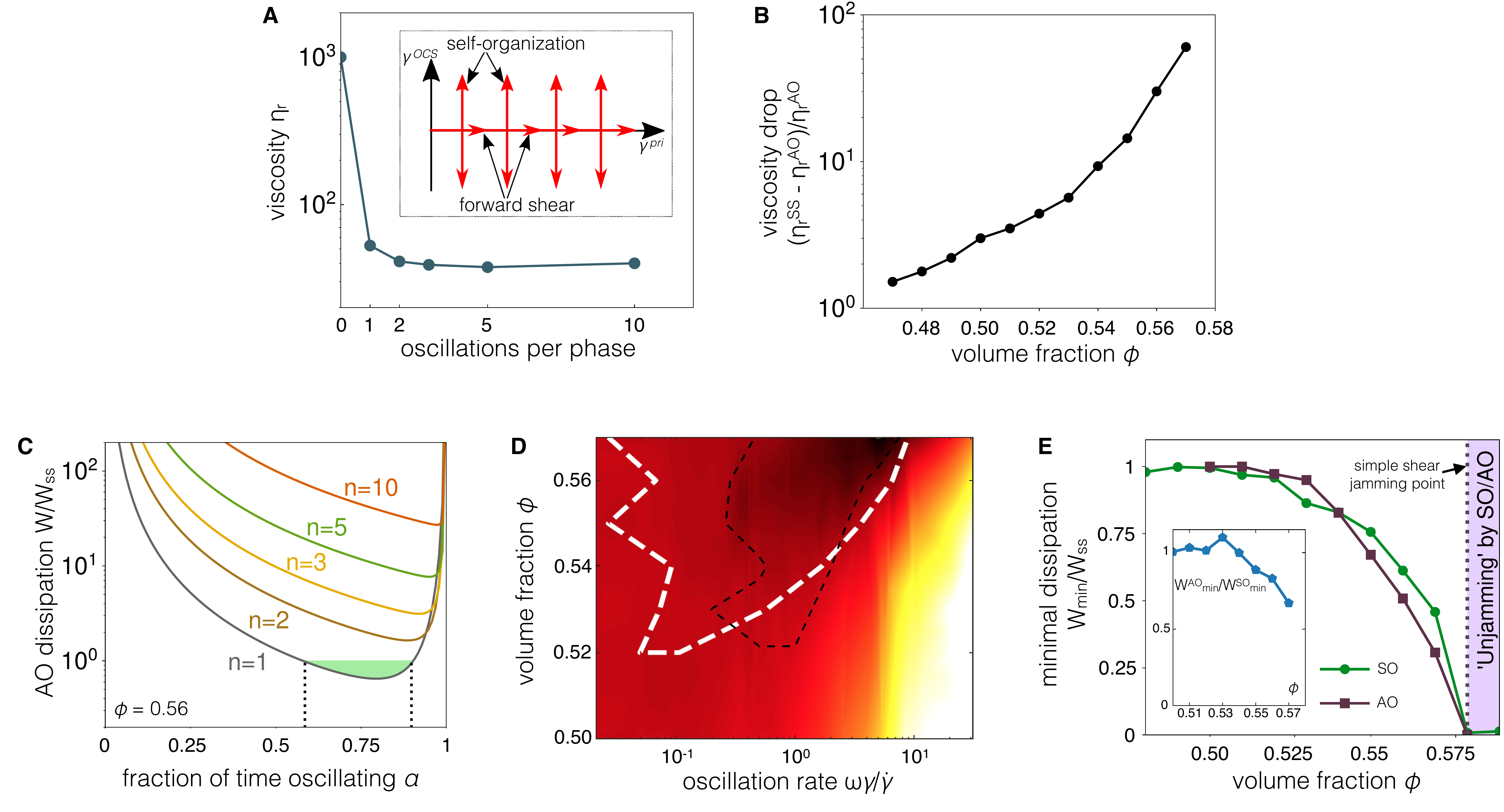}
 \caption{
  Optimizing dissipation reduction using an alternating \OSP{} (AO) protocol.
  \textbf{A}: Primary viscosity as a function of $n$, the number of cross shear oscillations per phase of the AO protocol,
  with $\OSPamp=\Gamma=\SI{1}{\percent}$ and $\phi=0.56$. Inset: schematic of the AO strain profile;
  \textbf{B}: Relative viscosity drop as a function of volume fraction $\phi$, comparing steady shear and AO;
  \textbf{C}: Energy dissipation for the cases $n=1,2, 5,10$ as a function of the fraction of time spent oscillating $\alpha$;
  \textbf{D}: Dissipation in the ($\OSPfreq\OSPamp/\SSrate$, $\phi$) plane rescaled by the $\phi$-dependent steady shear dissipation,
 highlighting the region for which dissipation is reduced by at least \SI{5}{\percent} with AO. Outlined in black is the analogous region from Fig~\ref{figure1}F[iii] for comparison;
  \textbf{E}: Comparison of the minimal dissipation obtained with the AO and SO protocols across a range of volume fractions $\phi$. Inset: the ratio between the minimal dissipation rates achievable for AO and SO, $W^\text{AO}_\text{min}$ and $W^\text{SO}_\text{min}$, respectively.
  }
 \label{figure3}
\end{figure*}

\subsec{Alternating OCS (AO): separated flow phases reduce dissipation further}
In the SO protocol above, particle contacts are eliminated by applying oscillatory cross shear concurrently
with the desired primary shear.
A relatively high energy cost arises from the need to have sufficiently fast oscillations to
ensure that random organization can compete with the
restoration of frictional particle contacts caused by the primary shear.
If this is indeed the mechanism, though, there is no strict requirement that we perform these flows concurrently.
Instead we can use alternating intervals of \OSP{} without primary shear and of
primary shear without \OSP{}. The former eliminates particle contacts; the latter restores them, but not before a finite strain has been achieved. The cross shear dissipation can in principle be reduced to zero by having long intervals of very slow oscillations, creating an optimization scenario different to that of SO.

We therefore now test a new flow protocol (``alternating \OSP'', AO) that alternates an interval of
 $n$ periods of oscillation during a time $\alpha T$ with
$\dot{\gamma}^\mathrm{pri}=0$ and ${\gamma}^\mathrm{\OSP}(t) = \OSPamp \sin(\OSPfreq t)$
for $\OSPfreq = 2\pi n/(\alpha T)$,
with an interval of primary shear during a time $(1-\alpha) T$ with
$\dot{\gamma}^{\mathrm{pri}}=\dot\gamma/(1-\alpha)$ and $\dot{\gamma}^\mathrm{\OSP}=0$.
Guided by our result in Fig~\ref{figure1}B, the oscillations are applied transverse to the primary flow,
where we anticipate that their efficacy will be maximized.
Averaged over one cycle $T$, the shear rate in the primary direction is $\dot\gamma$.
The primary shear strain during each cycle is $\Gamma$, i.e. $T = \Gamma/\dot\gamma$.

In the case of rate-independent dynamics as simulated above, the microstructure depends
on the strain path only (sketched in~\reffig{figure3}A [Inset]), not the rate at which it is followed.
As a consequence, the viscosity depends on $n$, $\OSPamp$ and $\Gamma$,
but not on $\alpha$ (as long as $0<\alpha<1$).
We define the relative viscosity as $\eta_r = \sigma_{xy}/(\eta\dot{\gamma}^{\mathrm{pri}})$.
This viscosity is averaged over the intervals of pure primary shear during the AO protocol, measured over
a time period covering 30 strain units in the primary direction. It is reported as a function of $n$ in~\reffig{figure3}A
for $\phi=0.56$ and $\OSPamp = \Gamma = \SI{1}{\percent}$.
The viscosity drops rapidly with $n$ and, remarkably, even $n=1$ is sufficient to achieve a viscosity reduction of $\approx96\%$ in this unjammed system.
Interestingly, this viscosity reduction is already larger than that achieved with SO, as seen by comparing the relative reductions in~\reffig{figure3}B (AO) and~\reffig{figure1}D [Inset] (SO).
Finally, we find that the viscosity drop is maximized with AO (as with SO~\cite{Lin2016a}) for
$\OSPamp,\ \Gamma \approx 1-5 \%$, while for larger $\OSPamp$
interparticle gaps close during cycles allowing frictional particle contacts and a rapid viscosity increase.
A viscosity transient for the AO protocol is given in the SI.

Although $\alpha$ has no role in setting the viscosity,
it is a crucial parameter when it comes to the dissipation, which depends on the deformation rate.
In~\reffig{figure3}C we show the work per unit primary strain $W$ (rescaled by the $\phi-$dependent steady shear dissipation $W_\mathrm{SS}$)
as a function of $\alpha$ for several values of $n$, for $\phi=0.56$.
For $n=1$ and for $\alpha$ values between 0.58 and 0.9,
the overall dissipation is reduced compared to steady shear,
reaching a reduction of around \SI{45}{\percent} at $\alpha\approx 0.8$.
Dissipation is minimized when $n=1$ for all $\phi$,
as the shear rate shoots up quickly with increasing $n$,
swamping any further viscosity reduction achieved for $n>1$.
To ease the comparison with the SO protocol,
we define an oscillatory frequency for AO as $\OSPfreq = 2\pi n\OSPamp/\Gamma$
(that is, a given frequency corresponds to the same number of cross shear oscillations per unit strain in the primary direction for SO and AO).
In~\reffig{figure3}D we show a map of the relative dissipation $W/W_\mathrm{SS}$
in the $(\phi, \OSPfreq)$ plane, showing a wider area of reduced dissipation with AO compared to SO (see~\reffig{figure1}F[iii]).
The AO protocol therefore has clear advantages particularly in avoiding the need to precisely tune the frequency of the driving oscillations.
We finally present in \reffig{figure3}E a comparison of the reduction in energy dissipation achieved
by the AO and SO protocols as a function of the volume fraction $\phi$.
For $\phi \lesssim 0.54$ there is no further gain with AO compared to SO,
but for larger volume fractions,
where the viscosity reduction performance of AO is markedly superior (Fig~\ref{figure3}B),
there is indeed an improvement in AO over SO.
We quantify this improvement in~\reffig{figure3}E [Inset], giving the ratio of the minimal dissipation for AO and SO ($W^\text{AO}_\text{min}/W^\text{SO}_\text{min}$). This shows that close to the steady shear jamming volume fraction
the improvement of AO over SO can reach almost \SI{40}{\percent}, suggesting that
AO is likely the protocol of choice for dissipation reduction in very concentrated frictional suspensions.

\section*{Concluding remarks}
Our results show that non-steady deformation protocols can lead
to substantial viscosity and energy dissipation reductions in any friction-dominated
suspension flow.
The strategy is applicable for most flows involving granular suspensions and related systems in which frictional particle contacts bear most of the stress in steady shear, including Brownian suspensions under very large stresses~\cite{mari2015discontinuous}.
Because of their simplicity, our protocols, or ones like them, might be readily implemented as \emph{precision} unblockers and flow controllers in industrial devices such as extruders or mixers, or as dissipation regulators for active granular damping~\cite{sack2013energy}.
In particular, an extrusion nozzle might be fitted with an internal coaxial cylindrical actuator that oscillates about its axis
with a protocol specified to maximize flowability according to our present results.
Moreover, such implementations might be applied not only to minimize viscosities,
but to regulate them against a desired set point.
We tested such a protocol numerically, with good success (see SI).
From a fundamental point of view, the relation to random organization opens new research directions. For example, it
suggests that protocols other
than oscillatory flow that lead
to a similar absorbing phase transition~\cite{tjhung_discontinuous_2016} might
also be good candidates for driven flow enhancement in complex fluids. It also suggests an unexpected link between rheological properties and hyperuniformity~\cite{hexner2015hyperuniformity,tjhung2015hyperuniform}.

\section*{Numerical method}

We simulate the trajectories of athermal, noninertial particles using a minimal model that comprises short-ranged hydrodynamic lubrication and frictional surface contacts. Our simulations comprise $\mathcal{O}(10^3)$ particles with size ratio $1:1.4$ in a periodic box.
For a particle pair with positions ${\bm x}_1$, ${\bm x}_2$ and translational and rotational velocities ${\bm U}_1$, ${\bm U}_2$ and ${\bm \Omega}_1$, ${\bm \Omega}_2$, respectively, in a background flow described at ${\bm x}_1$ by ${\bm U}^\infty({\bm x}_1) = {\bm E}^\infty{\bm x}_1 + {\bm \Omega}^\infty \times {\bm x}_1$, the hydrodynamic forces ${\bm F}^h_1$, ${\bm F}^h_2$ and torques ${\bm \Gamma}^h_1$, ${\bm \Gamma}^h_2$ are given by~\cite{jeffrey1984calculation,jeffrey1992calculation,kim1991microhydrodynamics}:
\begin{equation}
  \begin{bmatrix}
{\bm F}^h_1\\
{\bm F}^h_2\\
{\bm \Gamma}^h_1\\
{\bm \Gamma}^h_2\\
  \end{bmatrix}
  =
{\bm R}_\mathrm{Lub}
    \begin{bmatrix}
{\bm U}^\infty({\bm x}_1) - {\bm U}_1\\
{\bm U}^\infty({\bm x}_2) - {\bm U}_2\\
{\bm \Omega}^\infty - {\bm \Omega}_1\\
{\bm \Omega}^\infty - {\bm \Omega}_2\\
 {\bm E}^\infty\\
{\bm E}^\infty
  \end{bmatrix}
  +
  {\bm R}_\mathrm{Stokes}
      \begin{bmatrix}
{\bm U}^\infty({\bm x}_1) - {\bm U}_1\\
{\bm U}^\infty({\bm x}_2) - {\bm U}_2\\
{\bm \Omega}^\infty - {\bm \Omega}_1\\
{\bm \Omega}^\infty - {\bm \Omega}_2\\
  \end{bmatrix}
  \text{.}
\end{equation}
The matrices ${\bm R}_\mathrm{Lub}$ and ${\bm R}_\mathrm{Stokes}$ follow our earlier description~\cite{mari2014shear}, while the scalar resistances therein comprise only the leading short-ranged diverging contributions, following~Ref~\cite{ball1997simulation}. The hydrodynamic stress contribution for particle 1 resulting from its pairwise interaction with particle 2, with force ${\bm F}_1^h$ and particle-particle vector ${\bm r}$ is given by ${\bm S}^h = \frac{1}{2} ({\bm F}_1^h {\bm r}^\mathrm{T} + ({\bm F}_1^h)^\mathrm{T} {\bm r})$.

The leading terms of ${\bm R}_\mathrm{Lub}$ diverge according to $1/h$ as particles 1 and 2 approach, with $h$ the surface-surface distance. Following experimental evidence that lubrication layers break down in suspensions under large stress~\cite{fernandez2013microscopic}, and, equivalently, for large particles~\cite{Guy2015}, we use a minimum $h_\mathrm{min}=0.001a$ (with $a$ the smaller particle radius), below which hydrodynamic forces are regularised and particles may come into contact. For a particle pair with contact overlap $\delta$ and centre-centre unit vector ${\bm n}$, we compute the contact force and torque according to \cite{cundall1979discrete}:
\begin{subequations}
\begin{equation}
{\bm F}^c_1 = k_n\delta{\bm n} - k_t{\bm u}
\end{equation}
\begin{equation}
{\bm \Gamma}^c_1 = a_1 k_t({\bm n} \times {\bm u})
\end{equation}
\end{subequations}
where ${\bm u}$ represents the incremental tangential displacement, reset at the initiation of each contact. $k_n$ and $k_t$ are stiffnesses and $a_1$ is the radius of particle 1. The tangential force component is restricted by a Coulomb friction coefficient $\mu_s$ such that $|k_t{\bm u}| \leq \mu_s k_n\delta$. For larger values of $|k_t{\bm u}|$, contacts enter a sliding regime. The contact stress contribution is given by ${\bm S}^c = {\bm F}_1^c {\bm r}^\mathrm{T}$ for particle-particle vector ${\bm r}$ and pairwise force ${\bm F}_1^c$. The stress tensor is ${\bm \sigma} = 2\eta{\bm E}^\infty + \frac{1}{V}(\sum{\bm S}^h + \sum{\bm S}^c)$ where $\eta$ is the suspending fluid viscosity and the sums are over all relevant pairwise interactions. Throughout the main text, we focus on the shear component in the primary flow direction, $\sigma_{xy}$.

Trajectories are computed from the above forces using two equivalent schemes. In the first, contact and hydrodynamic forces and torques are summed on each particle~\cite{Ranga2017,cheal2018rheology} and the trajectory is updated according to Newtonian dynamics (using LAMMPS~\cite{plimpton1995fast}), ensuring the Stokes number ($\rho\dot{\gamma}a^2/\eta$ for particle density $\rho$, suspending fluid viscosity $\eta$ and shear rate $\dot{\gamma}$) remains $\ll1$ to approximate over-damped conditions. We also set $2\dot\gamma a/\sqrt{k_n/(2\rho a)} < 10^{-5}$ to approximate hard spheres. In the second, per-particle forces are explicitly set to zero and the velocities are computed to balance contact and hydrodynamic forces and torques, ensuring strictly inertia-free flow~\cite{seto_discontinuous_2013,mari2014shear}.
The numerical model generates results that are consistent with $\mu(J)$-rheology as predicted by the experimental work of~Ref~\cite{boyer2011unifying}, see SI.

\textbf{Acknowledgement} CN acknowledges an EPSRC Impact Acceleration Account and grant EP/N025318/1, and subsequently the Maudslay-Butler Research Fellowship at Pembroke College, Cambridge, for financial support and thanks Ranga Radhakrishnan for sharing his derivation of the hydrodynamic forces. MEC is funded by the Royal Society. Codes required to generate the results in this article are given at [\url{https://doi.org/10.17863/CAM.13416
}] and [\url{https://doi.org/10.5281/zenodo.999197}].

\bibliography{biaxial_ref}

\end{document}